\documentclass[prl,twocolumn,amsmath,amssymb,floatfix,showpacs,superscriptaddress]{revtex4}

\usepackage{bm}

\usepackage{amssymb,amsmath}
\usepackage{graphics}
\usepackage{subfigure}
\usepackage{epstopdf}
\usepackage{epsfig}
\usepackage{color}
\usepackage{amsfonts}
\usepackage{bm}
\usepackage{amscd}
\usepackage{epsfig}
\usepackage{subfigure}
\usepackage{color}
\usepackage{color}
\definecolor{darkgray}{rgb}{0.3,0.3,0.3}
\definecolor{gray}{rgb}{0.5,0.5,0.5}
\definecolor{yellow}{rgb}{.4,.4,0}
\definecolor{orange}{rgb}{1,0.5,0}
\definecolor{darkgreen}{rgb}{0,0.5,0}
\definecolor{darkblue}{rgb}{0,0,1}
\definecolor{darkred}{rgb}{0.5,0,0}
\definecolor{purple}{rgb}{0.35,0,0.35}\newcommand{\up}{\uparrow}
\newcommand{\down}{\downarrow}

\def\sig{{\mbox{\boldmath{$\sigma$}}}}

\begin{document}
\title{Suspended nanowires as mechanically-controlled Rashba spin-splitters
 }

\author{R. I. Shekhter}
\affiliation{Department of Physics, G\"{o}teborg University, SE-412 96 G\"{o}teborg, Sweden}

\author{ O. Entin-Wohlman}

\affiliation{Raymond and Beverly Sackler School of Physics and Astronomy, Tel Aviv University, Tel Aviv 69978, Israel}

\affiliation{Physics Department, Ben Gurion University,  Beer Sheva
84105, Israel}

\author{A. Aharony}


\affiliation{Raymond and Beverly Sackler School of Physics and Astronomy, Tel Aviv University, Tel Aviv 69978, Israel}
\affiliation{Physics Department, Ben Gurion University,  Beer Sheva
84105, Israel}\date{\today}

\begin{abstract}

Suspended  nanowires are shown to provide  mechanically-controlled coherent mixing/splitting of   the  spin states of transmitted electrons, 
caused by the Rashba spin-orbit interaction. The sensitivity of the latter
to mechanical bending makes the wire
a tunable nano-electro-mechanical  (NEM)  weak link between reservoirs. When the reservoirs are populatedÊ with misbalanced  ``spin up/down"  electrons, the wire becomes  a source of split spin currents, which are not associated with electric charge transfer and which do not depend on temperature or driving voltages. The mechanical vibrations of the bended wires allow for additional  tunability of these splitters by applying a magnetic field and varying the temperature. Clean metallic  carbon nanotubes of a few microns length 
are good candidates  for 
generating spin conductance  of the same order as the charge conductance (divided by $e^2$)  which would have been induced by electric driving voltages.

\end{abstract}

\pacs{07.10Cm,72.25.Hg,72.25.Rb}

\maketitle

\noindent{\bf Introduction.}  The lack of screening and   the wavy nature of the electrons  together with the ensuing  
interference effects
determine a large variety of Coulomb-correlation and quantum-coherence phenomena in quantum wires and  dots. The electronic spin, being weakly coupled to other degrees of freedom in bulk materials, becomes an ``active player"
due to the enhanced spin-orbit interaction induced by the Rashba effect \cite{Rashba} in these low-dimensional structures \cite{Theor_so_coup,Naaman}. This interaction can be also  modified experimentally  \cite{Kuemmeth,Flensberg,Jhang}.  The quantum-coherence control of
spin-related devices  and the spatial transfer of the electron spins
are among the most challenging tasks of nowadays spintronics, as they can bring up new functionalities. Thus, e.g.,
quantum interference of electronic waves in  multiply-connected  devices was predicted to be sensitive to the electronic spin, leading to spin filtering in electronic transport \cite{AA}.


In charge transport,
electronic beam splitters (e.g., by tunnel barriers) are  key ingredients in  interference-based devices. In this Letter we propose that tunnel-barrier scatterers may serve as  coherent splitters of the electronic spin when the tunneling 
electrons also undergo   spin (Rashba) scattering. This allows to map  various quantum-interference based phenomena   in charge transport onto  electronic spin transportation. Such spin-splitters can be readily made functional  by adding to them a mechanical degree of freedom, which serves to control  their geometrical configuration in space, to which the Rashba interaction is quite sensitive.   Due to this, one achieves   mechanical coherent control and mechanical tuning of the spin filters  \cite{comaa}.

\begin{figure}[htp]
\includegraphics[width=7cm]{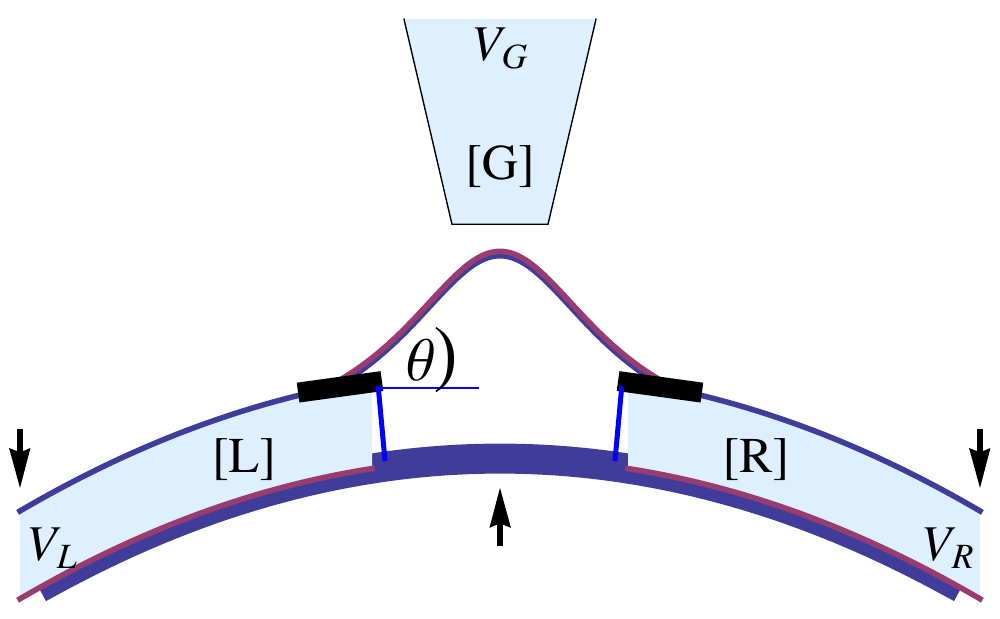}
\caption{A break junction supporting  a nanowire of length $d$
(possibly a carbon nanotube), 
attached by tunnel contacts  to
two biased electrodes ([L] and [R]). The small vibrations of the wire induce oscillations in the angle $\theta$ 
around some value $\theta_{0}$. The upper electrode ([G]) is an STM
tip biased differently. 
The Rashba interaction can be controlled via the bending angle $\theta$ of the wire. The latter  can be modified both mechanically,
by loads (shown by the arrows) applied to the substrate and electrically, by biasing the STM.
}
\label{fig1}
\end{figure}

We suggest that a suspended nanowire, acting as a weak link between two electronic reservoirs, is a good candidate for such a Rashba spin-splitter (see Fig.  \ref{fig1}).
The amount of spin splitting, brought about by the Rashba interaction on the wire,   is determined by the spin-orbit coupling and as such can be controlled 
by bending  the wire.
This
can be mechanically tuned, by
exploiting a break junction as a substrate for the wire
(see Fig. \ref{fig1}) or by electrically  inducing
a Coulomb interaction
between the wire and an STM tip
electrode (also displayed in Fig. \ref{fig1}).
This Rashba scatterer is localized on the nanowire, and  serves as a point-like scatterer in momentum-spin space for the electrons incident from the bulky leads. When 
there is a spin imbalance population in one of  the leads (or both), and the Rashba spin-splitter is activated (i.e., the weak link is open for electronic propagation) spin currents are generated and are injected from the point-like scatterer  to the leads. Thus the Rashba splitter redistributes the spin populations between the leads. This source of spin currents need not be accompanied by transfer of electronic charges \cite{combb}.



Such a coherent scatterer, whose scattering matrix can be ``designed" at will by tuning  controllably  the geometry,  
can be realized in electric weak-links based on clean carbon nanotubes (CNT). 
Carbon nanotubes have a significant Rashba spin-orbit coupling \cite{Theor_so_coup,Kuemmeth,Jhang}. Moreover, 
CNT's are known to have 
quite long mean-free paths (longer for {\em suspended } tubes that for the non-bended ones), 
allowing for experimental  detection of interference-based phenomena (e.g., Fabry-Perot interference patterns)
\cite{CNT}.


Further tunability of the Rashba spin-splitter can be achieved by switching on an external magnetic field,  coupled to the wire through the Aharonov-Bohm effect \cite{AB}. This is accomplished by  quantum-coherent displacements of the wire, which generate a  temperature dependence in the Aharonov-Bohm  magnetic flux (through an effective area) \cite{RIS}.
Generally,
 a large  mechanical deformability of  nanostructures, originating from their 
composite nature 
complemented by the  strong Coulomb forces accompanying single-electron charge transfer,  offer an additional functionality of electronic nanodevices \cite{hong,Blencowe}.  
Indeed,  coherent nano-vibrations  in suspended nanostructures,
with frequency in the gigahertz range,
 were detected experimentally \cite{OConnell}.



\begin{figure}[htp]
\includegraphics[width=8cm]{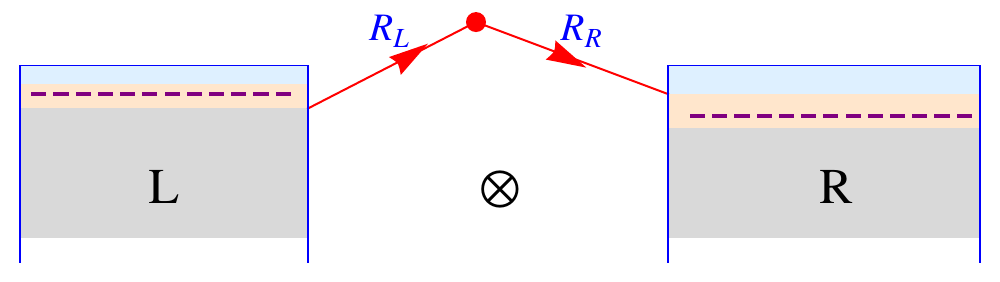}
\caption{Schematic geometry used for calculating the spin-orbit coupling dependence of the tunneling amplitude. A localized level 
is tunnel-coupled to left ($L$) and right ($R$) electronic electrodes with possibly different chemical potentials $\mu_{L\sigma}$ and $\mu_{R\sigma}$. The setup lies in the $x-y$ plane; a magnetic field applied along $\hat{\bf z}$   is shown by $\otimes$. The setup corresponds 
to a configuration in which the wire is controlled only mechanically, and the STM is not shown.}
\label{fig2}
\end{figure}

\noindent{\bf The transmission amplitude through a Rashba scatterer.} The model system exploited in the calculations is depicted in Fig. \ref{fig2}. There, the nanowire  is replaced by a quantum dot  (a widely-accepted picture, see Ref. \onlinecite{CNT}),  which has  a single level (of energy $\epsilon_{0}$), and which vibrates in the direction perpendicular to the wire in the junction plane. The  leads are modeled by free electron gases and  are firmly coupled to left and right reservoirs, of chemical potentials $\mu_{L\sigma}$ and $\mu_{R\sigma}$, respectively, allowing for spin-polarized charge carriers. Here, $\sigma$ denotes the spin index; the spin-quantization axis (assumed to be the same for both reservoirs)  depends on the spin imbalance in the reservoirs and will be specified below. 
The electronic populations in the reservoirs are thus
\begin{align}
f^{}_{L(R)\sigma}(\epsilon^{}_{k(p)})=[e^{\beta (\epsilon^{}_{k(p)}-\mu^{}_{L(R)\sigma )})}+1]^{-1}\ ,
\end{align}
with $\beta^{-1}=k_{\rm B}T$. The electron gas states in the left (right) lead are indexed by $k$ ($p$) and have energies $\epsilon_{k}$ ($\epsilon_{p}$). Below we denote
by $c_{k\sigma}$ ($c_{p\sigma}$) the annihilation operators for the leads, and  by  $c_{0\sigma}$ that for the localized level \cite{commc}.


\vspace{1.5cm}

\vspace{1cm}

The linear Rashba interaction  manifests itself as a phase factor on the tunneling amplitude
\cite{SOhop}. In the geometry of Fig. \ref{fig2}, this phase is induced by an electric field perpendicular to the $x-y$ plane,
and is given by
 $\alpha{\bf R}\times\sig\cdot\hat{\bf z}$, where     $\alpha$ denotes the strength of the spin-orbit interaction (in units of inverse length \cite{com}),   and $\sig$ is the vector of the Pauli matrices.
Quite generally ${\bf R}_{L}=\{x_{L},y_{L}\}$ for the left tunnel coupling  and   ${\bf R}_{R}=\{x_{R},-y_{R}\}$
for the right one, where both radius vectors ${\bf R}_{L}$ and ${\bf R}_{R}$ are functions of the vibrational degrees of freedom (as specified in the following).  The quantum vibrations of the wire  which modify the bending angle,  
make the electronic motion effectively two-dimensional.
This leads to the possibility of manipulating  the junction   via the Aharonov-Bohm effect, by applying a magnetic field  which imposes  a further  phase on the tunneling
amplitudes $\phi_{L(R)}=-(\pi/\Phi_{0})(Hx_{L(R)}y_{L(R)})$, where $H$ is the magnetic field and $\Phi_{0}$ is the flux quantum (a factor of order unity is absorbed in $H$ \cite{RIS}).



It follows that the tunneling Hamiltonian between the localized level and the leads takes the form
\begin{align}
{\cal H}^{}_{\rm tun}&=\sum_{k,\sigma,\sigma '}(V^{}_{k\sigma \sigma '}c^{\dagger}_{0\sigma}c^{}_{k\sigma '}
+{\rm H.c. })\nonumber\\
&+\sum_{p,\sigma,\sigma '}(V^{}_{p\sigma \sigma '}c^{\dagger}_{p\sigma}c^{}_{0\sigma '}+{\rm H.c. })\ .
\end{align}
The tunneling amplitudes are (operators  in spin and vibration spaces)
\begin{align}
V^{}_{k(p)}=-J^{}_{L(R)}\exp[-i\psi_{L(R)}]\ ,
\end{align}
where
\begin{align}
\psi^{}_{L}&=\phi^{}_{L}-\alpha(x^{}_{L}\sigma^{}_{y}-y^{}_{L}\sigma_{x}^{})\ ,\nonumber\\
\psi^{}_{R}&=\phi^{}_{R}-\alpha(x^{}_{R}\sigma^{}_{y}+y^{}_{R}\sigma_{x}^{})\ .\label{psi}
\end{align}
We consider a non-resonant case, where the localized level
is far above the energies of the occupied states in both leads  (i.e., no energy level on the wire is close enough to $\epsilon^{}_0$  
to be 
involved in  inelastic tunneling via a real state). This allows us to exploit the tunneling as an expansion parameter \cite{RIS} and to
preform a unitary transformation
which replaces the wire by an effective direct tunneling between the leads through virtual states
\begin{align}
{\cal H}^{\rm e}_{\rm tun}=
\sum_{k,p}(c^{\dagger}_{k}W^{\dagger}_{kp}c^{}_{p}+{\rm H.c.})\ ,
\end{align}
with  (using matrix notations in spin space)
\begin{align}
W^{\dagger}_{kp}=\frac{1}{2}\Bigl (\frac{1}{\epsilon^{}_{p}-\epsilon^{}_{0}}+\frac{1}{\epsilon^{}_{k}-\epsilon^{}_{0}}\Bigr )V^{\dagger}_{k}V^{\dagger}_{p}\ .\label{W}
\end{align}
A straightforward calculation \cite{SM} now yields that the spin-polarized particle flux emerging from the left lead is
\begin{widetext}
\begin{align}
I^{}_{L\sigma}=\int_{0}^{\infty}d\tau\sum_{k,p,\sigma '}\Bigl \{&f^{}_{R\sigma '}(\epsilon^{}_{p})[1-f^{}_{L\sigma}(\epsilon^{}_{k})]
\langle e^{i(\epsilon^{}_{k}-\epsilon^{}_{p})\tau} [W^{}_{pk}]^{}_{\sigma '\sigma}[W^{\dagger}_{kp}(\tau )]^{}_{\sigma\sigma'}+(\tau\rightarrow -\tau)\rangle\nonumber\\
&-f^{}_{L\sigma }(\epsilon^{}_{k})[1-f^{}_{R\sigma '}(\epsilon^{}_{p})]
\langle e^{i(\epsilon^{}_{p}-\epsilon^{}_{k})\tau} [W^{}_{kp}]^{}_{\sigma \sigma'}[W^{\dagger}_{pk}(\tau )]^{}_{\sigma'\sigma}+(\tau\rightarrow -\tau)\rangle\Bigr \}\ ,\label{IL}
\end{align}
\end{widetext}
where $\langle\rangle$ denotes thermal averaging over the vibrations and over the time evolution  with respect to the (free) Hamiltonians of the leads and the vibrations.
Assuming that the  $k$, $p$ dependence of the amplitudes may be ignored, and adopting the Einstein model for the description of the vibrations in the variable $\theta$ (see below), one readily obtains \cite{SM}
\begin{align}
&I^{}_{L\sigma}=\frac{\Gamma^{}_{L}\Gamma^{}_{R}}{2\pi\epsilon^{2}_{0}}\sum_{\sigma '}\sum_{n,n'=0}^{\infty}
P(n)|\langle n|[e^{-i\psi^{}_{R}}e^{-i\psi^{}_{L}}]^{}_{\sigma '\sigma}|n'\rangle|^{2}\nonumber\\
&\times(1-e^{\beta(\mu^{}_{L\sigma}-\mu^{}_{R\sigma '})})\frac{\mu^{}_{L\sigma}-\mu^{}_{R\sigma '}+(n'-n)\omega
}{e^{\beta[\mu_{L\sigma}-\mu_{R\sigma '}+(n'-n)\omega]}-1}\ ,\label{ils}
\end{align}
where $n$ is the vibrations' quantum number,  $P(n)=(1-\exp[-\beta\omega])\exp[-n\beta\omega]$,  and $\omega$ is the vibrations' frequency  ($\Gamma_{L(R)}$ are the usual partial widths induced on  $\epsilon_{0}$ by the coupling to the leads).
The  particle flux emerging from the right lead is obtained from Eq.  (\ref{IL}) upon interchanging the roles of the left and right sides of the junction, 
with $\sum_{\sigma}(I_{L\sigma}+I_{R\sigma})=0$, as required by charge conservation.
One notes [see Eqs. (\ref{psi})] that while the phase due to the magnetic field disappears in the absence of the vibrations,   
this is not so for the spin-orbit-phase (as $\psi_{L}$ and $\psi_{R}$ do not commute).

\noindent{\bf The  Rashba scatterer as a spin source.} Combining
 the expressions for the incoming spin fluxes [Eq. (\ref{ils}) and the corresponding one for $I_{R\sigma}$]  yields a net spin current, which is injected from the Rashba scatterer into the leads. Therefore, the scatterer can be viewed as a source of spin current, which is maintained when the leads have imbalanced populations. This spin current is defined as
\begin{align}
J^{}_{\rm spin}\equiv\sum_\sigma \sigma J^{}_{\rm spin,\sigma}=\sum_\sigma  \sigma(I^{}_{L\sigma}+I^{}_{R\sigma})\ , 
\end{align}
and it tends to diminish the spin imbalance in the leads, through spin-flip transitions induced by the Rashba interaction. In the limit of weak tunneling, 
we expect the spin imbalance to be kept constant in time by injecting 
spin-polarized electrons into the reservoirs, so that   
the (spin-dependent) chemical potentials  do not vary. 

 The explicit expressions for the two spin currents yield dramatic consequences  \cite{com1}: 1. Independent of the choice of the spin-quantization axis,  $J_{\rm spin,\sigma}$ is given solely by
the term with $\sigma '=\overline{\sigma}$ in the spin sums of Eq. (\ref{ils}) and the corresponding one for $I_{R\sigma}$ 
($\overline{\sigma}$ is the spin projection opposite to $\sigma$), which implies that only the off-diagonal (in spin space) amplitudes contribute. 2. Adopting the plausible geometry,  $y_{L}=y_{R}=(d/2)\sin(\theta)$ and $x_{L}=x_{R}=(d/2)\cos(\theta)$ where $d$ is the wire length and $\theta$, which vibrates around  $\theta_{0}$,  is defined in Fig. \ref{fig1}, one finds that 
\begin{align}
&e^{-i\psi^{}_{R}}e^{-i\psi^{}_{L}}=e^{i\frac{\pi Hd^{2}\sin (2\theta )}{4\Phi^{}_{0}}}\Bigl (1-2\cos^{2}(\theta )\sin^{2}(\alpha d/2)\nonumber\\
&+i\sigma^{}_{y}\sin (\alpha d)\cos (\theta )-i\sigma^{}_{z}\sin (2\theta )\sin^{2}(\alpha d/2)\Bigr )\label{10}
\ .
\end{align}
This result is independent of the choice
of the spin polarizations in the leads, and  {\em  does not involve $\sigma_{x}$}. 
3. As Eq. (\ref{10}) indicates, spin flip will be realized in our device
for any orientation
of the leads' polarization. Furthermore, if the angle $\theta$ vibrates about a non-zero average value $\theta^{}_0$ then both terms on the second line in Eq. (\ref{10}) yield spin flips even for the non-vibrating wire!  In this respect, the spin-orbit splitting effect is very different from that of the Aharonov-Bohm field, which requires a finite area and therefore in our setup is entirely caused by the mechanical vibrations.  In the special case $\theta^{}_{0}=0$, the second term there does not contribute for the non-vibrating wire, and then one has spin flips only if the polarization is in the $x-z$ plane.
To be concrete, below we present explicit results for a  
quantization axis  along $\hat{\bf z}$.

Let the chemical potentials of the two leads be
\begin{align}
\mu^{}_{L,R\up}=\mu^{}_{L,R}+\frac{U^{}_{L,R}}{2}\ ,\ \ \mu_{L,R\down}=\mu^{}_{L,R}-\frac{U^{}_{L,R}}{2}\ ,
\end{align}
such that the bias is given by $\Delta\mu =\mu_{L}-\mu_{R}$ while the amount of polarization in each of the leads is determined by $U_{L(R)}$.
Equation  (\ref{ils}) then yields
\begin{widetext}
\begin{align}
-J^{}_{{\rm spin},\down}=J^{}_{{\rm spin},\up}&=\sin^{2}(\alpha d)\frac{\Gamma^{}_{L}\Gamma^{}_{R}}{2\pi\epsilon^{2}_{0}}\sum_{n,n'}P(n)|\langle n|e^{i\frac{\pi H d^{2}}{4\Phi^{}_{0}}\sin (2\theta )}\cos (\theta )|n'\rangle |^{2}\nonumber\\
&\times \Bigl (\frac{[1-e^{\beta (\Delta\mu +U)}][\Delta\mu +U+(n'-n)\omega]}{e^{\beta [\Delta \mu +U+(n'-n)\omega ]}-1}
+
\frac{[e^{\beta (\Delta\mu -U)}-1][\Delta\mu -U+(n'-n)\omega]}{e^{\beta [\Delta \mu -U+(n'-n)\omega ]}-1}
\Bigr )\ ,\label{13}
\end{align}
\end{widetext}
where $U=(U^{}_{L}+U^{}_{R})/2$. Clearly, the spin intensity vanishes for non-polarized leads, for which $U=0$. One also observes that when the vibrations are ignored, i.e., when $\theta=\theta_{0}$
(and consequently only the $n=n'$ terms survives) the Aharonov-Bohm flux drops out, and
\begin{align}
J^{}_{\rm spin,\sigma}&=G^{}_{0}U\sin^{2}(\alpha d)\cos^{2}(\theta^{}_{0})
 \ ,
\end{align}
where $G^{}_{0}=\Gamma^{}_{L}\Gamma^{}_{R}/(\pi\epsilon^{2}_{0})$  is the Landauer zero-field electric conductance (divided by $e^2$, $e$ being the electronic charge).
Finally, 
when the spin intensity is linear in the chemical potentials (i.e., in the linear-response regime)
it loses its dependence on the bias voltage
and becomes
\begin{align}
J^{}_{{\rm spin},\up}=-UG^{}_{\rm spin}\ ,
\end{align}
with the ``spin conductance"
\begin{align}
&G^{}_{\rm spin}=G^{}_{0}
\sin^{2}(\alpha d)
\sum_{n=0}^{\infty}\sum_{\ell =1}^{\infty}P(n)\nonumber\\
&\times|\langle n|e^{i\frac{\pi H d^{2}}{4\Phi^{}_{0}}\sin (2\theta )}\cos (\theta )|n+\ell\rangle |^{2}\frac{2\ell \beta\omega}{e^{\beta\ell \omega}-1}\ .\label{Gspin}
\end{align}

Our final expression for the amount of spin intensity is obtained upon expanding  $\theta=\theta_{0}+\Delta\theta=\theta_{0}+(a^{}_{0}\cos (\theta_{0})/d)(b+b^{\dagger})$, where $b$ ($b^{\dagger}$) is the destruction (creation) operator of the vibrations, and $a_{0}$ is the amplitude of the zero-point oscillations.
Equation (\ref{Gspin})  then becomes
\begin{align}
&G^{}_{\rm spin}=\sin^{2}(\alpha d)\cos^{2}_{}(\theta^{}_{0})\nonumber\\
&\times\Bigr [G^{}_{ 0}\sum_{n=0}^{\infty}\sum_{\ell =1}^{\infty}|\langle n|e^{i\frac{H}{H^{}_{0}}(b+b^{\dagger})}|n+\ell\rangle |^{2}\frac{2P(n)\ell \beta\omega}{e^{\beta\ell \omega}-1}\Bigr ]\ ,\label{Gspin1}
\end{align}
with the magnetic-field scale given by $H_{0}=\sqrt{2}\Phi_{0}/[\pi da^{}_{0}\cos (\theta_{0})\cos (2\theta_{0})]$. Interestingly enough, the expression in the square brackets of Eq. (\ref{Gspin1}) is exactly the magnetoconductance (divided by $e^{2}$) of the wire, as analyzed in Ref. \onlinecite{RIS}.
We hence may build on their results to obtain for the spin-intensity admittance the low- and high-temperature limits
\begin{align}
\frac{G^{}_{\rm spin}/G^{}_{0}}{\sin^{2}(\alpha d)\cos^{2}_{}(\theta^{}_{0})}=\Biggl \{
\begin{array}{cc}1-\frac{\beta\omega}{6}\frac{H^{2}_{}}{H^{2}_{0}}\,& \beta\omega\ll1\\
\exp [-H^{2}/H^{2}_{0}] ,& \beta\omega\gg 1\ .
\end{array}
\end{align}

\noindent{\bf Discussion.}
In conclusion, we have proposed that electro-mechanically tunable interference of waves
of electronic spins
can be achieved in nanostructures with spatially localized spin-orbit interaction. Electric weak links
in a mechanically-controllable geometry
enable one
to exploit the Rashba spin-orbit interaction
to split electronic spins and to induce spin currents in polarized conductors. 
These currents are not associated with electric charge transportation. 
The Rashba spin-splitter is characterized by a scattering matrix that can be ``designed" at will, 
by mechanically tuning the nanowire.

Carbon nanotubes are in particular suitable for realizing the Rashba spin splitter. The energy gap induced  by the spin-orbit 
coupling in them is 0.37 meV, making the strength $\alpha $ on the order of $10^4$ cm$^{-1}$
\cite{Kuemmeth}.
For wire lengths of the order of micrometers, $\alpha d$ is of order unity, and then 
$ G_{\rm spin}$  is of the same order as the Landauer conductance (which determines the
response of electric currents to electric driving voltages).
It is then possible to tune the spin currents by an external electric field
(which controls the Rashba coupling).

Any experimental detection  of  spin patterns of electrons flowing through a nanotube would be an appropriate method to monitor the spin current injected from the Rashba splitter. Spin-dependent tunneling is one possibility. When the leads   are spin polarized,   the densities of states  at the Fermi energy of the left (right)  lead, ${\cal N}_{L(R)}^{\sigma}$,   depend
on the spin direction.  Then the electric conductance,  $G_{0}$,  becomes a function of the mechanical angle $\theta_{0}$ due to the Rashba-induced spin-flip transitions, leading to
\begin{align}
J^{}_{\rm spin}(\theta^{}_{0})
\propto U[G^{}_{0}(\theta^{}_{0})-G^{}_{0}(\frac{\pi}{2})]\Big |\frac{{\cal N}^{\sigma}_{L}{\cal N}^{\overline{\sigma}}_{R}-
{\cal N}^{\overline{\sigma}}_{L}{\cal N}^{\sigma}_{R}}{
{\cal N}^{\sigma}_{L}{\cal N}^{\overline{\sigma}}_{R}+
{\cal N}^{\overline{\sigma}}_{L}{\cal N}^{\sigma}_{R}}\Big |\ .\nonumber
\end{align}
Thus an electric measurement of $G_{0}(\theta_{0})$ can detect  the spin current. 
Another scheme which avoids a voltage drop is  to exploit the Rashba splitter as a superconducting weak link.   The full consideration of this setup is beyond the present  scope. However, having in mind the well-known $\pi-$shift in the Josephson current as a function of the superconducting phase difference \cite{Spivak} resulting from spin flips induced by impurities in the tunnel barrier, one may hope to observe mechanically-controlled phase shifts in the Josephson current arising from the Rashba splitting.

The possibility to electrically and mechanically activate point-like sources of spin-polarized currents with controllable orientations of the spin polarization opens a new route to study spin-related interference in more complicated arrays and networks. It couples such phenomena with electronic transport switching caused by e.g., Coulomb blockade or microwave activation.



\begin{acknowledgments}
We thank S. Matityahu for helpful discussions. The hospitality  of the
Kavli Institute for Theoretical Physics China, CAS, Beijing 100190, China,  where this work had begun, is gratefully acknowledged.
This work was supported by the Israeli Science Foundation (ISF) and the US-Israel Binational Science Foundation (BSF).
\end{acknowledgments}


\begin{thebibliography}{999}
\bibitem{commc}
The (small) effect of the Zeeman spin-splitting (caused when an external magnetic field is applied) on the density of states of the tunneling electrons is disregarded.
\bibitem{com1}
Because of charge-current conservation the total intensity, one has $\sum_{\sigma}\sigma  J^{}_{\rm spin ,\sigma}=2J_{\rm spin,\uparrow}=-2J_{\rm spin,\downarrow}$.
\end{thebibliography}
\end{document}